\documentclass[a4paper,fleqn]{cas-sc}

\usepackage[utf8]{inputenc}
\usepackage[english]{babel}
\usepackage{graphicx}
\usepackage{bm}

\usepackage{xcolor}
\usepackage{tcolorbox}

\usepackage[authoryear]{natbib}
\def\tsc#1{\csdef{#1}{\textsc{\lowercase{#1}}\xspace}}
\tsc{WGM}
\tsc{QE}
\ExplSyntaxOn
\keys_set:nn { stm / mktitle } { nologo }
\ExplSyntaxOff

\begin{document}
\let\WriteBookmarks\relax
\def\floatpagepagefraction{1}
\def\textpagefraction{.001}

\shorttitle{}
%\shortauthors{}

\title [mode = title]{Variational approach to determine the properties of dislocations at finite deformation}

\author[1,2]{István Groma}[orcid=0000-0002-6644-1365]
%\cormark[1]
\cortext[1]{Corresponding author}
\ead{groma@metal.elte.hu}
\credit{Conceptualization of this study, Theoretical foundation, Original draft preparation}
\affiliation[1]{organization={Institute for Solid State Physics and Optics, HUN-REN Wigner Research Centre for Physics},
            addressline={Konkoly T. út},
            city={Budapest},
            postcode={1525},
            state={},
            country={Hungary}}

\affiliation[2]{organization={Department of Materials Physics, E\"otv\"os Lor{\'a}nd University},
            addressline={Pázmány P. sétány. 1/A},
            city={Budapest},
            postcode={1117},
            state={},
            country={Hungary}}

\begin{abstract}
A generalised version of the continuum theory of curved dislocations describing the spatial and temporal  evolution of the fields: statistically stored dislocation density, geometrically necessary dislocation density, and curvature has recently been proposed.   The dynamics of the system are derived from a scalar functional of the relevant fields that cannot increase during the evolution of the system.
However, the framework was established only for small deformations. The details of the theory can be found in the PRB paper Ref.~\cite{groma2021dynamics}.

The aim of the present paper is to  discuss the fundamentals of the elasticity theory of finite deformation in cases where individual dislocations are present in the system. The equilibrium equations are derived within a variational formalism. It is shown that  introducing dislocations into the finite deformation framework is a nontrivial task. Moreover, if the deformation is large, the force acting on a dislocation segment is not the well-known Peach-Koehler force.  In a forthcoming paper, the results obtained will be applied to the generalisation of the dislocation continuum theory of curved dislocations.
\end{abstract}


%\item k
%\item kk
%\item kkk

\begin{keywords}
Dislocations \sep Finite deformation theory  \sep Variational approach \sep  Crystal plasticity
\end{keywords}

\maketitle


\section{Introduction}

The plastic deformation of crystalline materials is controlled by the collective motion of a large number of dislocations. The typical dislocation density in deformed metals is on the order of  $\rho \sim 10^{14}$ m$^{-2}$, i.e., the average spacing between dislocation lines is about 100 nm. Thus, a deformed micron-sized sample already contains a large number of strongly interacting dislocations. So, following the time evolution of the dislocation system through discrete dislocation dynamics (DDD) simulations is computationally rather demanding
\citep{kubin1992,ghoniem1999,devincre2001,devincre2002,bulatov2006computer,arsenlis2007enabling,akhondzadeh2020dislocation,fan2021strain,berta2025identifying}.
For most problems, however, an appropriate continuum model should be suitable, but due to the large degrees of freedom in the dislocation system, modelling the plastic deformation of crystalline materials in terms of dislocations requires handling the problem with statistical physics methods. The most challenging issue in developing a continuum theory is that traditional statistical physics methods cannot be directly applied because dislocation motion is strongly dissipative and dislocations are flexible lines.

During the past three decades, an important finding was that the plastic response of samples with characteristic dimensions on the order of 10 \textmu m or less depends on the sample size \citep{fleck1997advances}.  Despite several attempts to incorporate internal length scales into phenomenological continuum theories by adding strain-gradient terms to the stress \citep{zhu1997strain,aifantis1999,fleck,gurtin} there is not a commonly accepted solution. One can expect that an appropriate  continuum theory of dislocations could model this size effect. 

Another key feature observed is the formation of dislocation patterns during plastic deformation.  Since the early 1960s, several theoretical and numerical attempts have been suggested to model patterning \citep{holt1970,hansen1986, aifantis1985,pontes}. However, since they are not directly linked to the specific properties of individual dislocations, they are fundamentally phenomenological approaches.

Patterning also served as a significant  motivation for the development of DDD methods \citep{kubin1992,ghoniem1999,devincre2001,devincre2002}, but the simulations are computationally extremely expensive and poorly scalable. Thus, DDD is still limited to specific problems \citep{devincre2001,devincre2002,hussein2015}. Recently, El-Azab and coworkers \citep{xia2015computational,lin2020implementation} used a continuum formulation based on vector dislocation densities, which can capture the evolution of dislocation patterns. However, this pseudo-continuum variant of DDD is a numerical rather than a fully theoretical model of dislocation patterning.

By systematically coarse-graining the evolution equation of individual dislocations, a continuum theory has been developed over the past 25 years in a strongly simplified quasi two-dimensional system of straight, parallel edge dislocations
\citep{groma1997link,zaiser2001statistical,groma2003spatial,groma2007dynamics,mesarovic2010,dogge2015,groma2016dislocation,valdenaire2016density} that has been successfully compared to DDD simulations \citep{groma2003spatial,yefimov2004comparison,groma2006debye,ispanovity2020emergence}. By now, it can be considered a well-established theory for the 2D problem it addresses. Moreover, it has shown that the model can be formulated as a specific phase field theory \citep{groma2007dynamics,groma2010variational,groma2015scale,groma2016dislocation}.  The most important feature of the theory is that it predicts dislocation patterning, even though it was not ``designed'' for it \citep{groma2016dislocation,wu2018instability,ispanovity2020emergence}.

 Since, however, dislocations are moving curved flexible lines, an appropriate continuum theory should account for this.  The kinematic theory of the evolution of curved dislocations was developed by Hochrainer \emph{et al.} (see in \citep{hochrainer2007three,sandfeld2010numerical,hochrainer2014continuum,hochrainer2015multipole}). The kinematics were initially derived in a 2+1 dimensional space, containing the dislocation line direction as an independent variable. A multipole expansion of the theory results in a formulation in terms of alignment tensors \citep{groma2021dynamics}.

To obtain a closed theory,  appropriate assumptions about the dynamics of the dislocation system have to be added to the kinematic equations. Based on the analogy with the 2D case \citep{groma2016dislocation}, it is assumed that there is a scalar functional of the different fields, called ``plastic potential'', which cannot increase during the evolution of the system. This condition imposes a strong restriction on the possible forms of the velocity fields \citep{groma2021dynamics}.

However, a rather strong limitation of the theory presented in the paper Ref.~\cite{groma2021dynamics} is that  only small deformations are considered. Here, by applying a variational approach, we discuss the fundamentals of the theory of finite deformation when "individual" dislocations are present in the system. In the first part of the paper, the "defect free" nonlinear elasticity problem is briefly revisited by considering the functional derivative of the energy with respect to the deformation field.  As a next step, the equilibrium equation is derived if  plastic distortion is present in the system. In the last part, we discuss how one can introduce an individual dislocation into the nonlinear elastic continuum in a systematic manner. It is shown that the commonly applied form of the Peach-Koehler  force, which gives the force acting on a dislocation segment, has to be modified if we go beyond linear elasticity.   

The results obtained will serve as a key input for the continuum theory of dislocations in the finite deformation framework. 

\section{Applying functional derivation to get the equilibrium equations}
Let us start with the simple case where we consider $N$ atoms
positioned at $\vec{x}^i \ \ i=1..N$. Knowing the interaction energy 
$V(\vec{x}^1,...,\vec{x}^N)$, the equilibrium condition obviously reads as:
\begin{equation}
 -\frac{\partial V}{\partial x_i^p}=F_i^p=0,  \ \ \ p=1..N
\end{equation}
This is straightforward to generalise for a continuum. In this case, the state of the system is described by the function $\vec{x}(\vec{X})$, where $\vec{x}$ is the position of the point $\vec{X}$ after the deformation.  It should be noted that to introduce dislocations, we need a periodic crystalline structure as a starting reference frame. Therefore, when the time evolution of the system is considered, it is more relevant to apply a  Lagrangian description than an Eulerian one.

The elastic energy $E$ is a functional of $\vec{x}(\vec{X})$. (Dynamic elastic effects are excluded because, in most cases, they are not relevant to the plastic deformation we will discuss below.)  According to the principle of virtual work, the change in energy can be expressed as:
\begin{equation}
\delta W= E[\vec{x}(\vec{X})+\delta\vec{x}(\vec{X})]- E[\vec{x}]=-\int \vec{f}(\vec{x}(\vec{X}))\delta\vec{x}(\vec{X}) d^3\vec{X}=0 \label{eq.f}
\end{equation}
where $\delta\vec{x}(\vec{X})$ is a small perturbation function. Since the perturbation can be arbitrary, Eq. (\ref{eq.f}) follows that $\vec{f}(\vec{x}(\vec{X}))=0$. At the same time, according to the common definition of the functional derivative,
\begin{equation}
 \frac{\delta E}{\delta x_i}=-f_i(\vec{x}(\vec{X})).
\end{equation}
Thus, the equilibrium condition can be formulated as follows.
\begin{equation}
 \frac{\delta E}{\delta x_i}=0,
\end{equation}
that can always be calculated formally for any given energy functional.

In the following, we first consider a defect free system. For simplicity, we do not consider body forces (they could be added in a straightforward manner), and we take an infinite body (in this discussion, we are only interested  in the bulk equations). In this case, a rigid body translation cannot modify the internal energy. So, $E[\vec{x}]$ cannot change if a constant displacement is applied. One commonly applied  possibility to ensure this is to assume that the energy does not depend directly on $\vec{x}(\vec{X})$, only on its derivative. By introducing the deformation gradient
\begin{equation}
 F_{ij}=\frac{\partial x_i}{\partial X_j}
\end{equation}
we can state that $E$ is a functional of ${\bm F}$. By applying the rules of functional (variational) derivation, one gets that
\begin{equation}
 \frac{\delta E}{\delta x_i}=-\partial_j \frac{\delta E}{\delta F_{ij}}=0,
 \label{eq.F}
\end{equation}
where $\partial_j$ denotes the partial derivative with respect to $X_j$. The quantity 
\begin{equation}
   \sigma^{1PK}_{ij}=\frac{\delta E}{\delta F_{ij}}
\end{equation}
is commonly referred to as the 1st Piola–Kirchhoff stress tensor (see \cite{bonet1997nonlinear}).

Moreover, since a rigid body rotation also cannot modify the internal energy,  $E[\vec{x}]$ can depend only on the elastic deformation 
\begin{equation}
 \epsilon_{ij}=\frac{1}{2}\left(F^T_{ik}F_{kj}-\delta_{ij}  \right)
\end{equation}
where $T$ denotes the  transpose of the tensor ($F^T_{ij}=F_{ji}$). For an $O_{ij}$ rigid body rotation $F_{ij}=O_{ij}$. Since $O_{ji}O_{jk}=\delta_{ik}$ the deformation tensor vanishes for a rigid body rotation. Consequently,
\begin{equation}
 \frac{\delta E}{\delta F_{ij}}=\frac{\delta E}{\delta \epsilon_{kl}}
 \frac{d \epsilon_{kl}}{d F_{ij}}=\frac{\delta E}{\delta \epsilon_{kj}}F_{ik}=F_{ik}\sigma^{2PK}_{kj} \label{eq.PKII}
\end{equation}
where ${\bm \sigma}^{2PK}$ is called the 2nd Piola–Kirchhoff stress tensor (see \cite{bonet1997nonlinear}). Since ${\bm \epsilon}$ is a symmetric tensor, ${\bm \sigma}^{2PK}$ is also symmetric.

It is important to emphasise that the energy may depend on the appropriate derivatives of the deformation tensor ${ \bm \epsilon}$ (called nonlocal elasticity), but ${\bm \sigma}^{2PK}$ can always  be calculated by applying the formal rules of functional derivation. So, let us take an energy function in the form:
\begin{equation}
 E=\int e(\epsilon_{ij},\epsilon_ {ij,k})dV
\end{equation}
where $e(\epsilon_{ij},\epsilon_{ij,k})$ is the energy density and $\epsilon_ {ij,k}=\partial_k\epsilon_{ij}$.
In this case, Eq. (\ref{eq.F}) still holds, but 
\begin{equation}
  \frac{\delta E}{\delta F_{ij}}=F_{ik}\frac{\partial e}{\partial \epsilon_{kj}} -
  F_{ik}\partial_l \frac{\partial e}{\partial \epsilon_{kj,l}}
\end{equation}
As a result, in the equilibrium equation, higher derivatives of the displacement field can appear. This extension to "nonlocal" terms has been suggested by \cite{aifantis1992role, aifantis2003update}.  Among other things, such a theory can lead to a natural regularisation of the dislocation core. The details are discussed in Ref.~\cite{lazar2005nonsingular, po2014singularity, groma2010variational}.

If one considers a finite body, a surface term has to be added to the volume term introduced above. In this case, during the functional derivation, surface terms appear that, together with the variation of the surface energy term, lead to boundary conditions. The details are not discussed here.

It should be noted that the results obtained above are well known in deformation theory (see \cite{bookDill, dimitrienko2010nonlinear, mase2009continuum}). Our goal was only to demonstrate how the concept of functional derivation can be applied to obtain the equilibrium equations. The formalism that is used in several field theories, such as electrodynamics or general relativity,  can be very useful in non-trivial and nonlocal situations in elasticity theory. 

\section{Equilibrium equations in case of plastic deformation}
As is well known, if plastic deformation is involved, the final shape of the deformed body is reached in two steps: plastic deformation and elastic deformation. Thus, the deformation gradient $F_{ij}$ is decomposed into the product of a plastic transformation given by $F_{ij}^p$ and an elastic transformation $F_{ij}^e$:
\begin{equation}
  F_{ij}=F_{im}^e F_{mj}^p \label{eq.pe}
\end{equation}
However, internal forces are generated only during elastic deformation. Thus, the elastic energy depends only on the elastic deformation
\begin{equation}
 \epsilon_{ij}^e=\frac{1}{2}\left(F_{mi}^eF_{mj}^e-\delta_{ij}  \right)
 \label{eq.ee}
\end{equation}
From Eq. (\ref{eq.pe})
\begin{equation}
  F_{ij}^e=F_{io} F_{oj}^{-p} \label{eq.e}
\end{equation}
where ${\bm F}^{-p}$ denotes the inverse of ${\bm F}^{p}$. From Eqs. (\ref{eq.ee},\ref{eq.e})
\begin{equation}
 \epsilon_{ij}^e=\frac{1}{2}\left(F_{mo} F_{oi}^{-p} F_{mp}F_{pj}^{-p}-\delta_{ij}  \right)=\frac{1}{2}\left(F{io}^{-pT}C_{op}F_{pj}^{-p}-\delta_{ij}  \right)
 \label{eq.ee2}
\end{equation}
where
\begin{equation}
 C_{op}=F^T_{om}F_{mp}
\end{equation}
is the Cauchy–Green strain tensor.

Still in equilibrium, the functional derivative of the energy with respect to $x_i$ should vanish at fixed plastic deformation. After a short, straightforward calculation, one obtains (for details, see the Appendix) that
\begin{equation}
 -\frac{\delta E}{\delta xi}=\partial_j \frac{\delta E}{\delta F_{ij}}=\partial_j \left[
 \frac{\delta E}{\delta \epsilon_{kl}^e}\frac{d \epsilon_{kl}^e}{d F_{ij}}
 \right]=\partial_j \left[F_{ip} F_{pk}^{-p}\sigma_{kl}^{2PK} F_{lj}^{-pT}    \right] =\partial_j \left[F_{ip}\sigma_{pj}^{2PK*}\right]=0 \label{eq:eq_plast}
\end{equation}
As seen in the case of plastic deformation, in the equilibrium equation, the stress that is defined by the form
\begin{equation}
   \sigma_{ij}^{2PK}=\frac{\delta E}{\delta \epsilon_{ij}^e}
\end{equation}
has to be replaced with an effective transformed stress ${\bm \sigma}^{2PK*}={\bm F}^{-p}{\bm \sigma}^{2PK}{\bm F}^{-pT}$. As mentioned above, for simplicity, an infinite body has been considered. A finite body can be treated in a similar way as the defect free case.

One can find that from the general form of ${\bm \epsilon}^e$ given by Eq. (\ref{eq.ee2}) in the small deformation limit, the energy is:
 \begin{equation}
  E=\int \frac{1}{2}L_{ijkl}(\epsilon_{ij}-\epsilon^p_{ij})(\epsilon_{kl}-\epsilon^p_{kl}) dV
 \end{equation}
 where $L_{ijkl}$ is the elastic modulus tensor and $\epsilon^p_{il}=(F^p_{mi}F^p_{mj}-\delta_{ij})/2$. From this, the commonly applied equilibrium equation is certainly  recovered.
 
 For further consideration, let us calculate the functional derivative  of the energy with respect to $F_{ij}^{-p}$: 
\begin{equation}
 \frac{\delta E}{\delta F_{ij}^{-p}}=\frac{\delta E}{\delta \epsilon_{kl}^e}\frac{d \epsilon_{kl}^e}{d F_{ij}^{-p}}=
 F_{ki} F_{kp} F_{pl}^{-p} \sigma_{lj}^{2PK}=
 C_{ip} F_{pl}^{-p} \sigma_{lj}^{2PK}=
 F^T_{ip} F_{pl}^{e} \sigma_{lj}^{2PK} \label{eq:EF}
\end{equation}
(for details, see the Appendix.) For shortness, the notation
\begin{equation}
 \sigma^{f}_{ij}=F^T_{ip} F_{pl}^{e} \sigma_{lj}^{2PK} \label{eq:sigmaf}
\end{equation}
is introduced.
As will be proven below, this quantity determines the force acting on a dislocation segment. Again, in the small deformation limit $\sigma^{f}_{ij}=\sigma_{lj}^{2PK}$.

\section{Single dislocation loop}
\begin{figure}[pos=ht]
 \centerline{\includegraphics[angle=0,width=6cm]{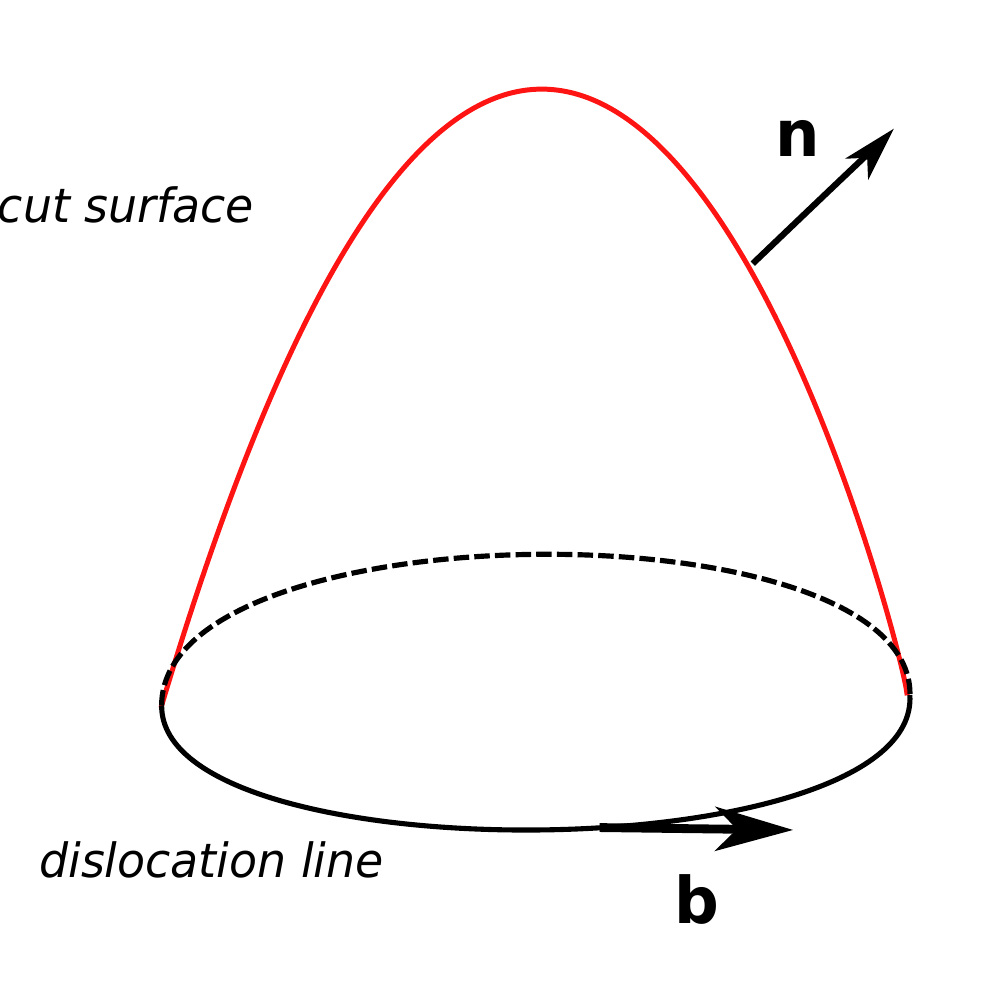}}
 \caption{Cut surface of a dislocation}
 \label{cutsurface}
\end{figure}

At the continuum level, a dislocation is defined by a cut surface on which the displacement field has a jump $\vec{b}$, called the Burgers vector (see Fig. \ref{cutsurface}).
As a consequence, the plastic deformation gradient reads as follows:
\begin{equation}
 F_{ij}^p=\delta_{ij}+b_i n_j \delta(\zeta)
\end{equation}
where  $\vec{n}$ is the normal vector of the cut surface at a given point on the surface, $\zeta$ is the distance from the surface in the $\vec{n}$ direction, and $\delta(x)$ is the Dirac delta function. The quantity
\begin{equation}
 \beta_{ij}^p=b_i n_j \delta(\zeta) \label{eq:beta}
\end{equation}
is called plastic distortion, i.e.
\begin{equation}
 F_{ij}^p=\delta_{ij}+\beta_{ij}^p \label{eq:Fp_s}
\end{equation}

In the energy expression, we need $F^e_{ij}$ as given by Eq. (\ref{eq.e}). So, one has to determine the inverse of $F^p_{ij}$ given above. Since Eq. (\ref{eq:beta}) contains a $\delta(x)$ function, obtaining the inverse of $F^p_{ij}$ is a nontrivial issue. Just to illustrate the problem, in a 1D version, the function
\begin{eqnarray}
 \frac{1}{1+b\delta(x)}
\end{eqnarray}
should be interpreted. Such a function does not have a clear mathematical meaning. However, the problem can be resolved if we approximate the Dirac delta with a regular function $\delta_b(x)$ with half width on the order of the Burgers vector (lattice parameter) $b$ and integral normalised to unity $\int_{-\infty}^{\infty} \delta_b(x)dx=1$. One possibility is a Gaussian function
\begin{eqnarray}
  \delta_b(x)\approx \frac{1}{sb\sqrt{\pi}}e^{-\frac{x^2}{s^2b^2}}
\end{eqnarray}
where $s$ is a parameter of the order of unity.
In this case, the maximum of $\delta_b(x)$ is of the order of $1/b$. Since in Eq. (\ref{eq:Fp_s}) we have $b \delta(x)$, at maximum, the approximate
$b \delta_b(x)$ is of the order of unity. This implies that
\begin{equation}
 F_{ij}^{-p}=\delta_{ij}-b_i n_j \delta(\zeta) \label{eq:F-p_s}
\end{equation}
is a good approximation up to linear terms in the Burgers vector.
The ``regularisation'' procedure is physically justified by the fact that, at the atomic level, the dislocation is defined by adding a half atomic plane to the crystal. Therefore,  the actual displacement ``jump'' is distributed within an interval with a length of the lattice parameter. As a result, at a continuum level, to avoid singularity problems  at the cut surface of a single dislocation, it is better to define $F_{ij}^{-p}$ as the primary quantity because only $F_{ij}^{-p}$ appears in the elastic energy.

Certainly, in the energy expression, one may also encounter the problem related to the singularity of $F^{-p}$. In order to avoid it, the ``regularised'' $\delta_b(x)$ function has to be used during the energy (or stress) calculation. If one takes the limit $s \rightarrow 0$, terms linear in the Burgers vector should only be kept. For higher orders in the Burgers vector, one has to specify  the dislocation core size that defines $s$.

%Another important issue is that in Eq. (\ref{eq:F-p_s}) the Burgers vector is obviously the ``deformed'' Burgers vector, i.e. $b_i=F^{ext}_{ij}b^{\rm lattice}_j$, where $b^{\rm lattice}_j$ is a primitive lattice vector in the undeformed reference crystal. Here $F^{ext}_{ij}$ denotes that part of the deformation gradient generated by the external load or other dislocations. With this, we exclude the self interaction of the dislocation. Moreover, one has to take into account that the cut surface also ``moves'' during the deformation, i.e. it is a function of $\vec{x}(\vec{X})$.

In a continuum theory of dislocations, one works with a coarse-grained  $F_{ij}^{-p}$. Thus, determining its inverse is not a problem, but to determine the force acting on a dislocation (see below), we have to consider the discrete problem. It should be noted that the issue of how to ``define'' $F^{-p}_{ij}$ for an individual dislocation raises the question of how to introduce the dislocation density tensor. According to \cite{kroner1981continuum}, the dislocation density is defined as
\begin{equation}
 \alpha_{ij}=e_{jkl}\partial_k \beta^p_{il}.
\end{equation}
With the form (\ref{eq:F-p_s}) suggested
\begin{equation}
\alpha_{ij}=-e_{jkl}\partial_k F^{-p}_{il}. \label{eq:alpha}
\end{equation}
However, based on different arguments, several alternative definitions are suggested (\cite{hochrainer2020crystal,acharya2001model,starkey2020theoretical}). Here  we mention the one proposed by \cite{gurtin, cermelli2001characterization}:
\begin{equation}
 \alpha_{ij}=\frac{1}{det(F^p)}F^{p}_{io}
 e_{jkl}\partial_k F^p_{ol}.
\end{equation}
However,  it requires the direct knowledge of $F^p_{ij}$, which, as explained above, is not straightforward to determine  for a single dislocation loop. The Eq. (\ref{eq:alpha}) suggested above is free from this problem.

It should also be noted that in phenomenological plasticity theories, the plastic velocity gradient
\begin{equation}
 L^p_{ij}=\dot{F}^p_{ik}F^{-p}_{kj}
\end{equation}
is often introduced and given as a sum of the rate of plastic deformation in the individual slip systems, but this is also somewhat problematic to define for an individual dislocation. In the continuum theory of curved dislocation proposed (\cite{groma2021dynamics}), this quantity is not needed.

\section{Force acting on a dislocation segment}

\begin{figure}[pos=ht]
 \centerline{\includegraphics[angle=0,width=6cm]
 {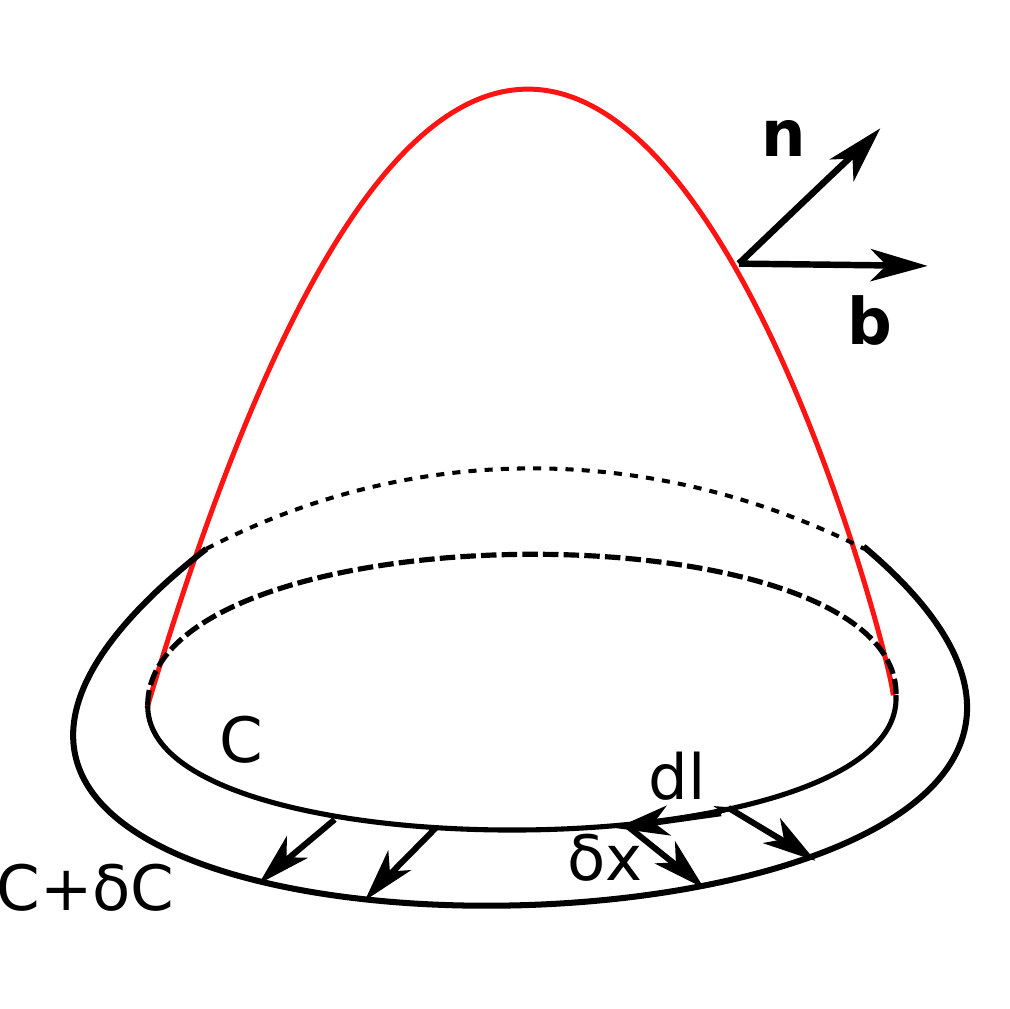}}
 \caption{Varying the dislocation line}
 \label{disl_line}
\end{figure}

A very important quantity is the force acting on a dislocation segment. For its calculation, let us move the dislocation line (see Fig. \ref{disl_line}).
Due to the motion of the dislocation line, $F^{-p}_{ij}$ varies, which is denoted by  $\overline{\delta} F^{-p}_{ij}$. What we have to calculate first is the energy change generated by this variation. Due to the line motion, the displacement field corresponding to the equilibrium configuration will also change. So, the total energy change is
\begin{equation}
 \overline{\delta}E=\int\left[ \frac{\delta E}{\delta F^{-p}_{ij}} \overline{\delta} F^{-p}_{ji}+
 \frac{\delta E}{\delta F_{ij}} \overline{\delta} F_{ji}
 \right] dV \label{eq:ee}
\end{equation}
where $\overline{\delta} F_{ji}$ denotes the variation of
$F_{ji}$ due to the dislocation line motion. Since 
$\overline{\delta} F_{ji}=\partial_i\overline{\delta} x_{j}$, where $\overline{\delta} x_{j}$ is the variation of the atomic position generated by moving the dislocation line,   with partial integration, the second term on the right hand side of Eq. (\ref{eq:ee}) can be rewritten as
\begin{equation}
 \overline{\delta}E=\int\left[ \frac{\delta E}{\delta F^{-p}_{ij}} \overline{\delta} F^{-p}_{ji}-
 \partial_i\left(\frac{\delta E}{\delta F_{ij}}\right) \overline{\delta} x_{j}
 \right] dV \label{eq:ee2}
\end{equation}
However, in the above equation, due to the equilibrium condition (\ref{eq:eq_plast}), the second term vanishes. Thus, the total energy change is
\begin{equation}
 \overline{\delta}E=\int\left[ \frac{\delta E}{\delta F^{-p}_{ij}} \overline{\delta} F^{-p}_{ji}
 \right] dV=\int \sigma^{f}_{ij}\overline{\delta} F^{-p}_{ji} dV \label{eq:ee3}
\end{equation}
where the notation introduced in (\ref{eq:sigmaf})
was used. It should be emphasised that during the calculation of $\sigma^{f}_{ij}$ one has to first calculate the functional derivative of the energy with respect to $F_{ij}$ at fixed $F^{-p}_{ij}$ and solve the
 equilibrium equation (\ref{eq:eq_plast}) for the displacement field. After that, the functional derivative of the energy with respect to $F^{-p}_{ij}$ must be calculated at fixed $F_{ij}$. Finally, the result has to be taken at $F_{ij}$ corresponding to the equilibrium solution.

Since $\overline{\delta} F^{-p}_{ij}$ is a delta function on the  stripe covered during the motion of the dislocation line, and the line displacement vector $\delta \vec{x}$ is small (see Fig. \ref{disl_line}), the integration can be reduced to a line integral along the dislocation line
\begin{equation}
\overline{\delta}E=-\oint \sigma^{f}_{ij} b_j
({\bf dl}\times {\bf \delta x})_i=-\oint ({\bm \sigma}^{f}\bf{b}
\times {\bf dl})){\bf \delta x}
\end{equation}
From this, the force acting on a unit length of a dislocation segment with line tangent $\bf{t}$
\begin{eqnarray}
  {\bf f}_{PK}=(\hat{\sigma}^{f} \bf{b}) \times \bf{t}
 \end{eqnarray}
should be considered the Peach-Koehler  force. As seen in large deformations, an effective stress appears in the Peach-Koehler  force. Certainly, in the small deformation limit, the common expression is recovered.

It should be noted that if we exclude climb and take the velocity law in the form 
\begin{eqnarray}
  { \bf f}_{||}=\lambda(|{\bf v}|) { \bf v}
 \end{eqnarray}
where $ {\bf f}_{||}$ is the projection of ${\bf f}_{PK}$ into the glide plane, and $\lambda(|\bf{v}|)$ is any positive function then
\begin{eqnarray}
\dot{E}= -\oint {\bf f}_{||} {\bf v} dl= -\oint\lambda(|{\bf v}|) {\bf v}^2 dl \leq 0.
 \end{eqnarray}
So, it is guaranteed that elastic energy cannot increase during the dislocation evolution; i.e., the dislocation motion is dissipative. 

An important consequence of the obtained results is that, to determine the force acting on the dislocation, the derivative of the energy functional must be calculated with respect to the plastic distortion at a fixed displacement field.
Certainly, to get the actual force, the displacement field has to be determined from the corresponding equilibrium equation. This result opens the way to generalise the continuum theory of dislocations, proposed earlier by (\cite{groma2021dynamics}), for large deformations. In a coarse-grained theory, where we operate with coarse-grained densities, the energy functional introduced above will play the role of the ``mean field''  part of the total phase field potential from which the continuum theory is derived. The $\hat{\sigma}^f$ ``mean field'' part of the stress is the functional derivative of the ``mean field'' energy functional. To get the total phase field functional, an appropriate ``correlation'' part has to be added (\cite{groma2021dynamics}). The generalised theory will be published in a forthcoming paper.

\section{Conclusions}
The field theory of individual dislocations is well established (see \cite{kroner1981continuum}). However, generalising it for a large deformation framework is not straightforward. It is shown that by applying a variational framework, similar to what is used in different field theories in physics, the appropriate equilibrium equations and the force acting on a dislocation segment can be derived in a systematic manner.  The method suggested allows one to go beyond linear and local elasticity. 

One may argue that linear elasticity is enough to determine the properties of individual dislocations. However, for a general theory of plasticity based on the collective evolution of a dislocation network, we have to handle the problem within a large deformation framework. According to the results obtained, in contrast to the small deformation level, the "stress measure" appearing in the equilibrium equation and in the Peach-Koehler force is not the same. The results obtained above will play the role of an "input" in the generalised version of the continuum theory of curved dislocations proposed in \cite{groma2021dynamics,groma2026dislocation}. 

\section{Appendix}
Here we provide the detailed derivation of some expressions used above.
If plastic distortion is present in the system, the elastic deformation is
\begin{equation}
 \epsilon_{kl}^e=\frac{1}{2}\left[F_{km}^{eT} F_{ml}^e-\delta_{kl}\right]= \frac{1}{2}\left[F_{mo}F_{ok}^{-p} F_{mp} F_{pl}^{-p}-\delta_{kl}
 \right]
\end{equation}
In the above considerations in Eq. (\ref{eq:eq_plast}) we need the derivative of $\epsilon_{kl}^e$ with respect to $F_{ij}$, that is:
\begin{equation}
 \frac{d  \epsilon_{kl}^e }{d F_{ij}}=\frac{1}{2}\left[
 \delta_{im}\delta_{jo}F_{ok}^{-p} F_{mp} F_{pl}^{-p}+
 F_{mo}F_{ok}^{-p}\delta_{mi}\delta_{jp} F_{pl}^{-p}  
 \right]=
\end{equation}
With this in Eq. (\ref{eq:eq_plast})
\begin{equation}
 \sigma_{kl}^{2PK}\frac{d  \epsilon_{kl}^e }{d F_{ij}}=\sigma_{kl}^{2PK}F_{jk}^{-p} F_{ip}F_{pl}^{-p}=F_{ip} F_{jk}^{-p} \sigma_{kl} F_{lp}^{-pT}=F_{ip}\sigma^{2PK*}_{pj}
\end{equation}

In Eq. (\ref{eq:EF}) we also need the derivative of $\epsilon_{kl}^e$ with respect to $F_{ij}^{-p}$:
\begin{equation}
 \frac{d  \epsilon_{kl}^e }{d F_{ij}^{-p}}=\frac{1}{2}\left[
 F_{mo}\delta_{io}\delta_{jk} F_{mp} F_{pl}^{-p}+
 F_{mo}F_{ok}^{-p} F_{mp} \delta_{ip}\delta_{jl} 
 \right]=
 \frac{1}{2}\left[
 \delta_{jk}F_{mi} F_{mp} F_{pl}^{-p}+
 F_{mo}F_{ok}^{-p} F_{mi} \delta_{jl} 
 \right]
\end{equation}
From this, we can see
\begin{equation}
  \sigma_{kl}^{2PK}\frac{d  \epsilon_{kl}^e }{d F_{ij}^{-p}}=
  \sigma_{kl}^{2PK}\delta_{jk}F_{mi} F_{mp} F_{pl}^{-p}=
  \sigma_{jl}^{2PK}F_{mi} F_{mp} F_{pl}^{-p}=
  C_{ip} F_{pl}^{-p}\sigma_{lj}^{2PK}
\end{equation}

\section*{Acknowledgement}
The financial support of the Hungarian National Research, Development and Innovation Office (IG and PDI, Project No. NKFIH EXCELLENCE25 153976) is acknowledged.

\bibliographystyle{cas-model2-names}

%\bibliography{citations}

\end{document}